\documentclass[a4paper]{article}

\usepackage{INTERSPEECH2022}
\usepackage{multirow}
\usepackage{booktabs}
\usepackage{array}
\usepackage{footmisc}

\title{A Comparative Study of Fusion Methods for SASV Challenge 2022}
\name{Petr Grinberg$^1$, Vladislav Shikhov$^1$}
\address{
  $^1$Samsung R\&D Institute Russia}
\email{ p.grinberg@samsung.com, v.shikhov@samsung.com}

\begin{document}

\maketitle
\begin{abstract}
Automatic Speaker Verification (ASV) system is a type of biometric authentication. It can be attacked by an intruder, who falsifies data in order to get access to protected information. Countermeasures (CM) are special algorithms that detect these spoofing-attacks.  While the ASVspoof Challenge series were focused on the development of CM for fixed ASV system, the new Spoofing Aware Speaker Verification (SASV) Challenge organizers believe that best results can be achieved if CM and ASV systems are optimized jointly \cite{jung2022sasv}. 
	
One of the approaches for cooperative optimization is a fusion over embeddings or scores obtained from ASV and CM models. The baselines of SASV Challenge 2022 present two types of fusion: score-sum and back-end ensemble with a 3-layer MLP. This paper describes our research of other fusion methods, including boosting over embeddings, which has not been used in anti-spoofing studies before.
\end{abstract}
\noindent\textbf{Index Terms}: SASV Challenge 2022, Countermeasure, Automatic Speaker Verification, Fusion

\section{Introduction}

Automatic Speaker Verification (ASV) system is a biometric authentication method working with humans' speech. As any type of protection it may be susceptible to tampering, called spoofing-attacks, which can be classified into four kinds: impersonation, voice conversion (VC), replay attacks and text-to-speech (TTS)~\cite{wu2015spoofing}. There are two ways, how to deal with spoofing-attacks. First is to improve the reliability of ASV systems. Second -- to use specialized algorithms, called countermeasures (CM), which detect unauthorized access.

The ASVspoof initiative and the first ASVspoof Challenge 2015 were created in order to standardize research in the field, metrics and databases~\cite{wu2015asvspoof}. This competition was held four times in 2015, 2017, 2019 and 2021 respectively. Last time it contained three tasks~\cite{delgado2021asvspoof}:
\begin{itemize}
	\item Logical Access (LA) task: spoofing data is created by VC and TTS algorithms and transmitted over telephone and VoIP networks.
	\item Physical Access (PA) task: CMs for replay attacks.
	\item Speech Deepfake (DF) task: combination of other tasks, but without speaker verification. 
\end{itemize}

The new Spoofing Aware Speaker Verification (SASV) Challenge 2022 is a continuation of the ASVspoof series. However, in ASVspoof CMs work with a fixed ASV system. Organizers of the SASV Challenge believe that joint optimization of CM and ASV systems can lead to more robust models. This paper compares different approaches for solution for this challenge.

Existing architectures for CM systems can be divided into two types: ones, which work with raw signal as input, and ones, which apply time-frequency transforms and create spectrograms. In \cite{tak2020explainability} authors showed that spoofing artifacts lie in different frequency sub-bands rather than in full-band. Because of that, the performance of spectrogram-based methods relies on the used time-frequency algorithm's resolution in the sub-band, where spoofing attack left marks. Different spectrograms enhance different frequencies, what leads to their contrasting capabilities. Hence, combining architectures with different front-ends can result in a robust model. Raw-input solutions define which frequencies are helpful by themselves during the train stage. However, their aptitude is still limited. Thus, architecture with great discriminating power should use fusion over models with different input types.

Fusion methods appeared in literature are usually done over scores obtained from different CM models. It was shown that non-linear types of fusion perform better than the linear ones~\cite{tak2020spoofing}. However, average or weighted sum with pre-normalization are still the most common~\cite{jung2022sasv, wang2021comparative, caceres2021biometric, chen2021pindrop, cai2017countermeasures}. Other appeared methods, where scores from CMs are stacked in one feature-vector, are presented below:
	
\begin{itemize}
	\item Support Vector Machines (SVM) with different type of kernels: linear, residual basis function (RBF) and polynomial. It was used in \cite{tak2020spoofing, tak2021end, kang2021crim}.
	
	\item Logistic Regression was tested by authors of \cite{tak2020spoofing, kang2021crim, chen2021ur}.
	
	\item Gaussian Mixture Model (GMM) appeared in \cite{tak2020spoofing}.
	
	\item MLP, Decision Tree and Random Forest were used as a final classifier over future-vectors in \cite{kang2021crim}.
\end{itemize}

All of these methods, except the Decision Tree, are shown to improve the performance and surpass single model systems. 

Fusion over embeddings occurs much less often. Authors of ~\cite{tak2021graph} created Fusion-Layer, which combines outputs of model sub-parts. SASV Challenge 2022 organizers used a 3-layer MLP with embeddings from both ASV and CM systems~\cite{jung2022sasv}.

In this paper we explore and compare the performance of different fusion methods. Apart from existing solutions, which mostly use scores, we take embeddings from CM and ASV systems. Moreover, only one ensemble method of Decision Trees was applied in anti-spoofing research -- bagging (specifically, Random Forest). We propose usage of another Decision Tree ensemble method called boosting. Concretely, fusion is done using CatBoost~\cite{prokhorenkova2018catboost}, and it significantly outperforms SASV Challenge 2022 baselines on all metrics. In addition, we test usage of Random Fourier Features (RFF)~\cite{rahimi2007random} with a logistic regression base, which is a RBF SVM approximation method.

The remainder of this paper is organized as follows. Section 2 describes CM and ASV systems, which embedding will be taken for fusion. Section 3 describes the experimental setup, which results are described in Section 4. Conclusions are presented in Section 5.

\section{CM and ASV systems}

\begin{figure*}[t]
	\centering
	\includegraphics[width=0.84\linewidth]{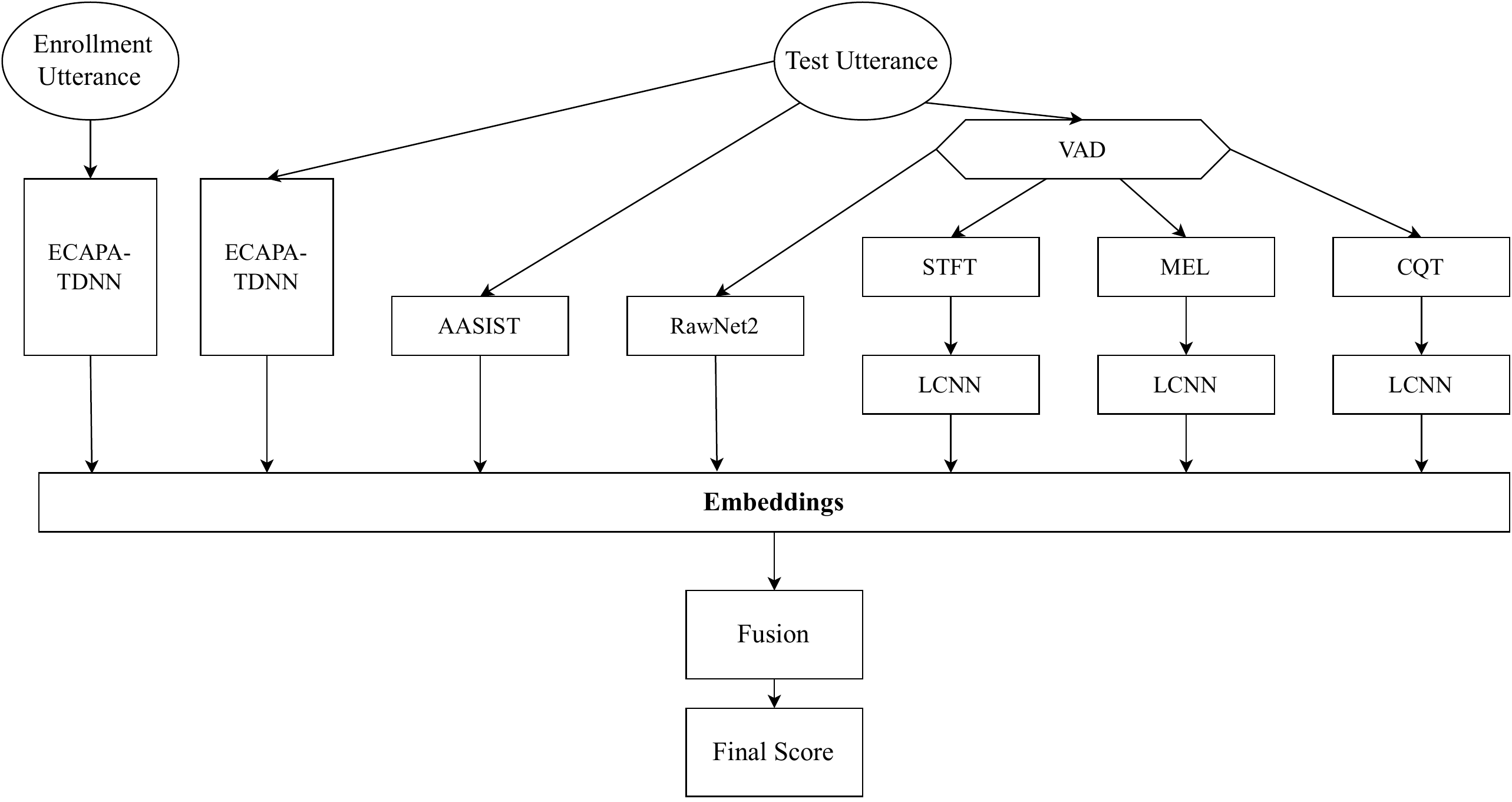}
	\caption{Pipeline for fusion over embeddings experiment}
	\label{fig:ExperimentalSetup}
\end{figure*}

This section describes what CM ans ASV systems were used in experiments.

\subsection{Countermeasure}

For the CM system we took five different models: two with raw input and three with spectrogram front-ends.

Firstly, we used the AASIST~\cite{jung2021aasist}, which is a baseline CM subsystem in the SASV Challenge 2022. This is a spectro-temporal graph attention network working with raw input. Sending waveform through AASIST results in 190-dimensional embedding. Scores are outputs of the last Fully-Connected layer (using softmax is an option).

Secondly, we adopted the RawNet2~\cite{tak2021end} architecture for our training pipeline. Specifically, Dropout layers were added. RawNet2 was a baseline CM system for the ASVspoof Challenge 2021~\cite{delgado2021asvspoof}. It consists of a Sinc-layer, ResBlocks and a GRU-layer, ending with two FC-layers. Embeddings are 1024-dimensional vectors obtained from the output of the first classifying FC-layer. Scores come out the last layer. 

Thirdly, we took the LightCNN (LCNN) architecture proposed in \cite{wu2018light} with three types of input, as was done in \cite{lavrentyeva2019stc}: STFT, MEL and CQT spectrograms. These time-frequency transform methods enhance different frequency sub-bands, which allows to create a robust model by combining them. LCNN is a classic CNN model with Max-Feature-Map operations as activation functions.  Embeddings are outputs of the penultimate FC-layer and have 80 dimensions for each type. Scores come out the last layer.

\subsection{Automatic Speaker Verification}

For the ASV system we used ECAPA-TDNN proposed in \cite{desplanques2020ecapa}, which is also a baseline ASV subsystem in SASV Challenge 2022. This subsystem consists of Squeeze-Excitation Res2Blocks, Statistic Pooling and Multi-layer Feature Aggregation. ECAPA-TDNN returns two 160-dimensional vectors for enrollment and test utterances respectively. Score is a cosine similarity between embeddings.

\section{Experimental Setup}

\begin{table*}[t]
	\caption{Results of Fusion Methods over embedding on Dev and Eval sets. Best value in column is bold. SASV Challenge 2022 Baselines' results and ECAPA-TDNN without any CM are at the bottom of the table. All EERs are in percents.}
	\label{tab:fusion_methods_embeddings_results}
	\centering
	\begin{tabular}{lcccccc}
		\toprule
		\multirow{3}{*}{\textbf{Fusion Method}}      & \multicolumn{6}{c}{\textbf{Metric}} \\
		\cmidrule{2-7}
		&  \multicolumn{2}{c}{SV-EER} & \multicolumn{2}{c}{SPF-EER} & \multicolumn{2}{c}{SASV-EER}\\
		\cmidrule{2-7}
		& Dev & Eval & Dev & Eval & Dev & Eval\\
		\midrule
		3-layer MLP & $48.58$ & $49.08$ & \boldmath$0.06$ & $0.85$ & $15.88$ & $25.10$\\
		Logistic Regression & $50.13$ & $50.03$ & $0.17$ & $1.50$ & $16.59$ & $25.41$\\
		SVM (Linear Kernel) & $49.88$ & $49.33$ & $9.13$ & $16.07$ & $17.70$ & $27.67$\\
		SVM (RBF Kernel) & $27.02$ & $22.99$ & $0.26$ & $1.32$ & $11.00$ & $12.24$\\
		SVM (Polynomial Kernel) & $29.14$ & $31.76$ & $7.41$ & $10.70$ & $13.92$ & $19.39$\\
		RFF (Logistic Regression Based) & $37.53$ & $36.24$ & $2.76$ & $8.73$ & $14.11$ & $21.08$\\
		GMM & $50.56$ & $49.95$ & $49.90$ & $49.47$ & $50.04$ & $49.64$\\
		Random Forest & $35.42$ & $32.73$ & $0.06$ & $0.76$ & $13.13$ & $18.16$\\
		\textbf{CatBoost} & $3.62$ & $3.87$ & \boldmath$0.06$ & \boldmath$0.61$ & \boldmath$2.09$ & \boldmath$2.90$\\
		\midrule
		Score-Sum (Baseline 1) & $32.88$ & $35.32$ & \boldmath$0.06$ & $0.67$ & $13.07$ & $19.31$\\
		3-layer MLP (Baseline 2) & $12.87$ & $11.48$ & $0.13$ & $0.78$ & $4.85$ & $6.37$\\
		\midrule
		ECAPA-TDNN & \boldmath$1.88$ & \boldmath$1.63$ & $20.30$ & $30.75$ & $17.38$ & $23.83$\\
		\bottomrule
	\end{tabular}
\end{table*}

In this section we describe how we trained our models and used fusion over them to following evaluation.

\subsection{Training Stage}

Our ASV system ECAPA-TDNN was pre-trained on the VoxCeleb2 dataset~\cite{chung2018voxceleb2}. It is same to the SASV Challenge 2022's organizers' ASV subsystem.

All CMs were trained on the ASVspoof 2019 LA train partition \cite{wang2020asvspoof}. We also took pre-trained AASIST as was done in competition. However, RawNet2 and LCNN-based models were trained from scratch and had data preprocessing with data augmentation.

In the DF task of the ASVspoof challenge speech waveforms were encoded and then decoded with different lossy codecs. This added some distortion into data and complicated the detection of spoofing-attacks. However, in practice compressed audio is common. So we believe that robust CM, which can be used in real-life situations, should handle these encoding-decoding perturbations. For this we added random compression with mp3 and aac codecs as data augmentation. 

There is a recent study, which shows that the ASVspoof Challenge Database has a very uneven distribution of silence: the bonafide samples have much longer silences, than many of attacks, especially TTS ones~\cite{muller2021speech}. This leads to a serious problem: models partially make their predictions based on the duration of silence in the data. It was shown that same architectures learned on samples with trimmed silence get much worse results. We believe that strong biometric authentication system has to deal with no silence data and reject samples with hush: indeed, if somebody wants access to protected information, he or she should provide his or her biometry, and there is no biometry in silence. Therefore, we preprocess data for RawNet2 and LCNN-based models by removing silences using simple magnitude-based Voice Activity Detection (VAD) algorithm to improve their detecting spoofing artifacts capabilities.

We trained RawNet2 and LCNN-based models for 30 epochs with the Adam optimizer.

\subsection{Fusion methods over embeddings}
We already discussed why fusion of CMs' scores with different inputs is better than the single system. We hypothesize that at the same manner fusion of CMs with and without silence trimming results in a well-performing model, which detects artifacts and does not make predictions based only on silence duration.

For this study we get embeddings from each sub-model and concatenate them into one long feature-vector, which is used as input for the final classifier. Then the classifier is trained on embeddings from train data. RawNet2 and LCNN-based models' embeddings are taken from preprocessed and encoded-decoded data. For each LCNN-based model input spectrogram is split into three equal-sized parts. We get embeddings for each part and stack them together.  

Final scores for development and evaluation sets are received from the trained classifier. We again used same preprocessing and encoding-decoding for RawNet2 and LCNN-based models. 

The resulting pipeline of our system is given in Figure \ref{fig:ExperimentalSetup}. The final feature-vector is a 2288-dimensional vector of numerical features.

Methods tested as final classifier are as follows:
\begin{itemize}
	\item 3-layer MLP. Same architecture, that was used in SASV Challenge 2022 Baseline, but with our 2288-dimensional vector as input: three FC-layers with 256, 128 and 64 out features respectively, LeakyReLU with $0.3$ negative slope as activation and one final classifier FC-layer.
	\item Logistic Regression. Iterations are limited to $1000$ (we also tried bigger value, but its performance was worse). Regular coefficient is $\lambda = \frac{1}{m}$, where $m=25380$ is the amount of utterances in ASVspoof 2019 LA train partition. All other parameters are set to default\footnote{\label{note1}We used sklearn v.0.24.2 \cite{scikit-learn}}.
	\item SVM with a Linear kernel. Iterations are limited to $50000$, regular coefficient is $\lambda = \frac{1}{m}$. All other parameters are set to default\footref{note1}.
	\item SVM with a RBF kernel. Iterations are limited to $50000$, regular coefficient is $\lambda = \frac{1}{m}$. All other parameters are set to default\footref{note1}.
	\item SVM with a Polynomial kernel. Iterations are limited to $50000$, regular coefficient is $\lambda = \frac{1}{m}$, degrees are set to $7$. All other parameters are set to default\footref{note1}.
	\item RFF with a logistic regression. Classifier uses linear principal component analysis (PCA) with $1024$ dimensions and Standard Scaler. Number of random features is set to $5000$. Iterations of inner logistic regression are limited to $50000$, regular coefficient is $\lambda = \frac{1}{m}$. All other parameters are set to default\footref{note1}.
	\item GMM. Number of components is set to $2$. Amount of EM iterations is $1000$. All other parameters are set to default\footref{note1}.
	\item Random Forest. Number of estimators is $1000$, all other parameters are set to default\footref{note1}.
	\item CatBoost. Number of estimators is $700$, all other parameters are set to default\footnote{We used catboost v.1.0.4 \cite{prokhorenkova2018catboost}}.
\end{itemize}

\subsection{Fusion methods over scores}
For completeness of our study we tested if we could improve the results by doing fusion over scores. First of all, we replicated SASV Challenge 2022 Baseline 2 (it will be called RB2 for shortness). Then we took RB2 and our best model (OBM for shortness) from the embeddings experiment and stacked their scores for each utterance in one 4-dimensional feature vector.

Then we tested the same classifiers, except MLP and RFF, because of too small input size. For Logistic Regression and SVM-based methods iterations were unlimited and regular coefficient was set to $10000$.

\section{Results}

\begin{table*}[t]
	\caption{Results of Fusion Methods over scores on Dev and Eval sets. Best value in column is bold. RB2, OBM, SASV Challenge 2022 Baselines' results and ECAPA-TDNN without any CM are at the bottom of the table. All EERs are in percents.}
	\label{tab:fusion_methods_scores_results}
	\centering
	\begin{tabular}{lcccccc}
		\toprule
		\multirow{3}{*}{\textbf{Fusion Method}}      & \multicolumn{6}{c}{\textbf{Metric}} \\
		\cmidrule{2-7}
		&  \multicolumn{2}{c}{SV-EER} & \multicolumn{2}{c}{SPF-EER} & \multicolumn{2}{c}{SASV-EER}\\
		\cmidrule{2-7}
		& Dev & Eval & Dev & Eval & Dev & Eval\\
		\midrule
		Logistic Regression & $3.81$ & $3.57$ & \boldmath$0.06$ & \boldmath$0.52$ & $2.20$ & \boldmath$2.75$\\
		SVM (Linear Kernel) & $3.81$ & $3.57$ & \boldmath$0.06$ & \boldmath$0.52$ & $2.20$ & \boldmath$2.75$\\
		SVM (RBF Kernel) & $3.90$ & $3.69$ & \boldmath$0.06$ & \boldmath$0.52$ & $2.15$ & $2.79$\\
		SVM (Polynomial Kernel) & $3.79$ & $3.54$ & \boldmath$0.06$ & $0.53$ & $2.14$ & $2.77$\\
		GMM & $8.19$ & $6.75$ & $2.49$ & $2.19$ & $5.53$ & $5.29$\\
		Random Forest & $7.42$ & $5.37$ & $1.19$ & $0.87$ & $3.63$ & $3.39$\\
		CatBoost & $6.73$ & $4.87$ & $0.19$ & $0.68$ & $2.83$ & $2.88$ \\
		\midrule
		RB2 & $13.38$ & $11.97$ & \boldmath$0.06$ & $0.68$ & $6.12$ & $6.25$ \\
		OBM & $3.62$ & $3.87$ & \boldmath$0.06$ & $0.61$ & \boldmath$2.09$ & $2.90$\\
		\midrule
		Score-Sum (Baseline 1) & $32.88$ & $35.32$ & \boldmath$0.06$ & $0.67$ & $13.07$ & $19.31$\\
		3-layer MLP (Baseline 2) & $12.87$ & $11.48$ & $0.13$ & $0.78$ & $4.85$ & $6.37$\\
		\midrule
		ECAPA-TDNN & \boldmath$1.88$ & \boldmath$1.63$ & $20.30$ & $30.75$ & $17.38$ & $23.83$\\
		\bottomrule
	\end{tabular}
\end{table*}

The results for different fusion methods over embeddings described in Section 3.2 are shown in terms of SV-EER, SPF-EER and SASV-EER in Table \ref{tab:fusion_methods_embeddings_results}. The performance of SASV Challenge 2022 Baseline systems and ECAPA-TDNN without any CM are at the bottom of the table. SV-EER measures how the model distinguish target and not-target trials. SPF-EER -- target and spoofed trials. SASV-EER is a general metric, where non-target and spoofed trials are treated equally.

We can clearly see that the GMM approach works as a random binary classifier. It was expected, because embeddings are hardly similar to Gaussian vectors. Hence, GMM should not be used for fusion over embeddings. 

Unlike the results from \cite{tak2020spoofing} where Polynomial SVM outperformed RBF and Linear ones, in our experiment SVM with RBF Kernel shows the best of three results. Logistic regression has similar results for RBF SVM and Poly SVM in terms of SPF-EER and SASV-EER respectively, but SV-EER is equal to random classification. Linear SVM's quality indicators are poor too. The reason for such performance of Logistic Regression and Linear SVM is the linear inseparability of bonafide and non-target trials, which can be observed by the superiority of non-linear approaches. 3-layer MLP has the same problem, caused by the small train set and narrow layers with large input size. In theory, Random Fourier Features should approximate SVM with a RBF kernel. We can see from Table \ref{tab:fusion_methods_embeddings_results} that RFF indeed has similar performance to SVM methods, but it has much worse results than RBF SVM.

Random Forest, which is a bagging ensemble method of Decision Trees, exceeds the results of Baseline 2 in terms of SPF-EER and has similar results to Polynomial SVM in other metrics. Thus, Random Forest is better than SVM with a polynomial kernel in our study. 

The final tested method is CatBoost. It surpasses all competitors methods in all metrics with a huge margin in SV-EER and SASV-EER. Moreover, it reduces error-rates, obtained by baselines, by a relative $71.8\%, 66.28\%, 0\%, 8.95\%, 56.90\%, 54.47\%$ for SV-EER, SPF-EER and SASV-EER on development and evaluation sets respectively. Only single ECAPA-TDNN outperforms the CatBoost approach on SV-EER. It is not surprising because using countermeasures worsens verification ability.

To further improve the results we took model with the CatBoost fusion method (OBM) and combined it with RB2 in our second experiment, described in Section 3.3. Its results are presented in Table \ref{tab:fusion_methods_scores_results}.

Scores are more likely to be from Gaussian Mixture than embeddings. The GMM approach still deteriorates performance of the best single model, however it is far from random classification. Hence, GMM is not hopeless as score fusion method.

Logistic Regression and SVM-based methods have alike results. Polynomial SVM has the best of four performances with only a slight impairment in terms of SPF-EER and SASV-EER on evaluation set. These four methods outperform OBM on the evaluation set and slightly worsen on the development set. Hence, these methods work as some kind of regularization.

Both the Decision Tree ensemble methods got unsatisfactory quality indicators, however, they are better than GMM.

Thus, our best results are obtained using SVM with a 7-degree polynomial kernel score fusion of RB2 and OBM, which is a fusion of ECAPA-TDNN, AASIST, RawNet2 and three LCNN models' embeddings using CatBoost. This system is our submission for the SASV Challenge 2022.

\section{Conclusions}

This paper reports a comparison of different fusion methods over embeddings. Results show that the CatBoost approach, which did not appear in anti-spoofing studies before, outperforms all other methods by a huge margin. Moreover, we trimmed silence in audio data for some of our sub-models, what was demonstrated to worse performance in \cite{muller2021speech}, but still have desirable results. This indicates how robust the CatBoost method is. Other fusion over embeddings approaches can still get satisfactory results, but only for the distinction of target and spoofing-trials. They have poor results on SV-EER and SASV-EER metrics.

However, for fusion over scores everything is different. Logistic Regression and SVM-based methods outperform the best single system on the evaluation set and act as a regularization. Random Forest and CatBoost did not get great results. Thus, this methods are better to use with embeddings than scores. Finally, GMM should be used only as a score fusion approach, and its performance highly depends on the distribution of scores from sub-models. In addition, our study confirms that non-linear fusion methods are better than the linear ones. This paper is submitted to INTERSPEECH 2022.

\bibliographystyle{IEEEtran}

\bibliography{A_Comparative_Study_of_Fusion_Methods_for_SASV_Challenge_2022}

\begin{thebibliography}{10}
\providecommand{\url}[1]{#1}
\csname url@samestyle\endcsname
\providecommand{\newblock}{\relax}
\providecommand{\bibinfo}[2]{#2}
\providecommand{\BIBentrySTDinterwordspacing}{\spaceskip=0pt\relax}
\providecommand{\BIBentryALTinterwordstretchfactor}{4}
\providecommand{\BIBentryALTinterwordspacing}{\spaceskip=\fontdimen2\font plus
\BIBentryALTinterwordstretchfactor\fontdimen3\font minus
  \fontdimen4\font\relax}
\providecommand{\BIBforeignlanguage}[2]{{%
\expandafter\ifx\csname l@#1\endcsname\relax
\typeout{** WARNING: IEEEtran.bst: No hyphenation pattern has been}%
\typeout{** loaded for the language `#1'. Using the pattern for}%
\typeout{** the default language instead.}%
\else
\language=\csname l@#1\endcsname
\fi
#2}}
\providecommand{\BIBdecl}{\relax}
\BIBdecl

\bibitem{jung2022sasv}
J.-w. Jung, H.~Tak, H.-j. Shim, H.-S. Heo, B.-J. Lee, S.-W. Chung, H.-G. Kang,
  H.-J. Yu, N.~Evans, and T.~Kinnunen, ``{SASV Challenge 2022: A Spoofing Aware
  Speaker Verification Challenge Evaluation Plan},'' \emph{arXiv preprint
  arXiv:2201.10283}, 2022.

\bibitem{wu2015spoofing}
Z.~Wu, N.~Evans, T.~Kinnunen, J.~Yamagishi, F.~Alegre, and H.~Li, ``{Spoofing
  and countermeasures for speaker verification: A survey},'' \emph{Speech
  Communication}, vol.~66, pp. 130--153, 2015.

\bibitem{wu2015asvspoof}
Z.~Wu, T.~Kinnunen, N.~Evans, J.~Yamagishi, C.~Hanil{\c{c}}i, M.~Sahidullah,
  and A.~Sizov, ``{ASVspoof 2015: the first automatic speaker verification
  spoofing and countermeasures challenge},'' in \emph{Sixteenth annual
  conference of the international speech communication association}, 2015.

\bibitem{delgado2021asvspoof}
H.~Delgado, N.~Evans, T.~Kinnunen, K.~A. Lee, X.~Liu, A.~Nautsch, J.~Patino,
  M.~Sahidullah, M.~Todisco, X.~Wang \emph{et~al.}, ``{ASVspoof 2021: Automatic
  speaker verification spoofing and countermeasures challenge evaluation
  plan},'' \emph{arXiv preprint arXiv:2109.00535}, 2021.

\bibitem{tak2020explainability}
H.~Tak, J.~Patino, A.~Nautsch, N.~Evans, and M.~Todisco, ``{An Explainability
  Study of the Constant Q Cepstral Coefficient Spoofing Countermeasure for
  Automatic Speaker Verification},'' in \emph{Proc. The Speaker and Language
  Recognition Workshop (Odyssey 2020)}, 2020, pp. 333--340.

\bibitem{tak2020spoofing}
------, ``{Spoofing Attack Detection Using the Non-Linear Fusion of Sub-Band
  Classifiers},'' in \emph{Proc. Interspeech 2020}, 2020, pp. 1106--1110.

\bibitem{wang2021comparative}
X.~Wang and J.~Yamagishi, ``{A Comparative Study on Recent Neural Spoofing
  Countermeasures for Synthetic Speech Detection},'' in \emph{Proc. Interspeech
  2021}, 2021, pp. 4259--4263.

\bibitem{caceres2021biometric}
J.~C{\'a}ceres, R.~Font, T.~Grau, J.~Molina, and B.~V. SL, ``{The Biometric Vox
  system for the ASVspoof 2021 challenge},'' in \emph{Proc. ASVspoof2021
  Workshop}, 2021.

\bibitem{chen2021pindrop}
T.~Chen, E.~Khoury, K.~Phatak, and G.~Sivaraman, ``{Pindrop Labs’ Submission
  to the ASVspoof 2021 Challenge},'' in \emph{Proc. ASVspoof 2021 Workshop},
  2021.

\bibitem{cai2017countermeasures}
W.~Cai, D.~Cai, W.~Liu, G.~Li, and M.~Li, ``{Countermeasures for Automatic
  Speaker Verification Replay Spoofing Attack : On Data Augmentation, Feature
  Representation, Classification and Fusion},'' in \emph{Proc. Interspeech
  2017}, 2017, pp. 17--21.

\bibitem{tak2021end}
H.~Tak, J.~Patino, M.~Todisco, A.~Nautsch, N.~Evans, and A.~Larcher,
  ``{End-to-end anti-spoofing with RawNet2},'' in \emph{ICASSP 2021-2021 IEEE
  International Conference on Acoustics, Speech and Signal Processing
  (ICASSP)}.\hskip 1em plus 0.5em minus 0.4em\relax IEEE, 2021, pp. 6369--6373.

\bibitem{kang2021crim}
W.~H. Kang, J.~Alam, and A.~Fathan, ``{CRIM’s system description for the
  ASVspoof 2021 Challenge},'' in \emph{Proc. ASVspoof 2021 Workshop}, 2021.

\bibitem{chen2021ur}
X.~Chen, Y.~Zhang, G.~Zhu, and Z.~Duan, ``{UR channel-robust synthetic speech
  detection system for ASVspoof 2021},'' in \emph{Proc. 2021 Edition of the
  Automatic Speaker Verification and Spoofing Countermeasures Challenge}, 2021,
  pp. 75--82.

\bibitem{tak2021graph}
H.~Tak, J.-w. Jung, J.~Patino, M.~Kamble, M.~Todisco, and N.~Evans,
  ``{End-to-end spectro-temporal graph attention networks for speaker
  verification anti-spoofing and speech deepfake detection},'' in \emph{Proc.
  2021 Edition of the Automatic Speaker Verification and Spoofing
  Countermeasures Challenge}, 2021, pp. 1--8.

\bibitem{prokhorenkova2018catboost}
L.~Prokhorenkova, G.~Gusev, A.~Vorobev, A.~V. Dorogush, and A.~Gulin,
  ``{CatBoost: unbiased boosting with categorical features},'' \emph{Advances
  in neural information processing systems}, vol.~31, 2018.

\bibitem{rahimi2007random}
A.~Rahimi and B.~Recht, ``{Random Features for Large-Scale Kernel Machines},''
  in \emph{Proceedings of the 20th International Conference on Neural
  Information Processing Systems}, ser. NIPS'07.\hskip 1em plus 0.5em minus
  0.4em\relax Curran Associates Inc., 2007, p. 1177–1184.

\bibitem{jung2021aasist}
J.-w. Jung, H.-S. Heo, H.~Tak, H.-j. Shim, J.~S. Chung, B.-J. Lee, H.-J. Yu,
  and N.~Evans, ``{AASIST: Audio Anti-Spoofing using Integrated
  Spectro-Temporal Graph Attention Networks},'' \emph{arXiv preprint
  arXiv:2110.01200}, 2021.

\bibitem{wu2018light}
X.~Wu, R.~He, Z.~Sun, and T.~Tan, ``{A Light CNN for Deep Face Representation
  with Noisy Labels},'' \emph{IEEE Transactions on Information Forensics and
  Security}, vol.~13, no.~11, pp. 2884--2896, 2018.

\bibitem{lavrentyeva2019stc}
G.~Lavrentyeva, S.~Novoselov, A.~Tseren, M.~Volkova, A.~Gorlanov, and
  A.~Kozlov, ``{STC Antispoofing Systems for the ASVspoof2019 Challenge},'' in
  \emph{Proc. Interspeech 2019}, 2019, pp. 1033--1037.

\bibitem{desplanques2020ecapa}
B.~Desplanques, J.~Thienpondt, and K.~Demuynck, ``{ECAPA-TDNN: Emphasized
  Channel Attention, Propagation and Aggregation in TDNN Based Speaker
  Verification},'' in \emph{Proc. Interspeech 2020}, 2020, pp. 3830--3834.

\bibitem{chung2018voxceleb2}
J.~S. Chung, A.~Nagrani, and A.~Zisserman, ``{VoxCeleb2: Deep Speaker
  Recognition},'' in \emph{Proc. Interspeech 2018}, 2018, pp. 1086--1090.

\bibitem{wang2020asvspoof}
X.~Wang, J.~Yamagishi, M.~Todisco, H.~Delgado, A.~Nautsch, N.~Evans,
  M.~Sahidullah, V.~Vestman, T.~Kinnunen, K.~A. Lee \emph{et~al.}, ``{ASVspoof
  2019: A large-scale public database of synthesized, converted and replayed
  speech},'' \emph{Computer Speech \& Language}, vol.~64, p. 101114, 2020.

\bibitem{muller2021speech}
N.~M. M{\"u}ller, F.~Dieckmann, P.~Czempin, R.~Canals, K.~B{\"o}ttinger, and
  J.~Williams, ``{Speech is Silver, Silence is Golden: What do ASVspoof-trained
  Models Really Learn?}'' in \emph{Proc. 2021 Edition of the Automatic Speaker
  Verification and Spoofing Countermeasures Challenge}, 2021, pp. 55--60.

\bibitem{scikit-learn}
F.~Pedregosa, G.~Varoquaux, A.~Gramfort, V.~Michel, B.~Thirion, O.~Grisel,
  M.~Blondel, P.~Prettenhofer, R.~Weiss, V.~Dubourg, J.~Vanderplas, A.~Passos,
  D.~Cournapeau, M.~Brucher, M.~Perrot, and E.~Duchesnay, ``{Scikit-learn:
  Machine Learning in {P}ython},'' \emph{Journal of Machine Learning Research},
  vol.~12, pp. 2825--2830, 2011.

\end{thebibliography}

\end{document}